# Generating large-scale network analyses of scientific landscapes in seconds using Dimensions on Google BigQuery.

Michele Pasin[*], Richard J. Abdill[**]

[*]*m.pasin@digital-science.com*, corresponding author
Digital Science, 6 Briset St, London, EC1M 5NR, UK

[**]*richard.abdill@pennmedicine.upenn.edu*
Digital Science, 6 Briset St, London, EC1M 5NR, UK
Current address: University of Pennsylvania, 3400 Civic Center Boulevard, Philadelphia PA, 19104

**Abstract**

The growth of large, programatically accessible bibliometrics databases presents new opportunities for complex analyses of publication metadata. In addition to providing a wealth of information about authors and institutions, databases such as those provided by Dimensions also provide conceptual information and links to entities such as grants, funders and patents. However, data is not the only challenge in evaluating patterns in scholarly work: These large datasets can be challenging to integrate, particularly for those unfamiliar with the complex schemas necessary for accommodating such heterogeneous information, and those most comfortable with data mining may not be as experienced in data visualisation. Here, we present an open-source Python library that streamlines the process accessing and diagramming subsets of the Dimensions on Google BigQuery database and demonstrate its use on the freely available Dimensions COVID-19 dataset. We are optimistic that this tool will expand access to this valuable information by streamlining what would otherwise be multiple complex technical tasks, enabling more researchers to examine patterns in research focus and collaboration over time.

**Introduction**

Across even the most disparate fields, meta-research is a key requirement for evaluating and improving the scientific enterprise. Understanding trends and gaps in how research is funded, performed and published allows data-driven introspection from funders, institutions, companies and individual researchers—what is being researched, and by whom? Which collaborations are most productive, and in what ways? Endless answers are available from comprehensive databases such as those provided by Dimensions, Crossref, Scopus and Web of Science (Thelwall 2018), for those who have the technical capabilities of asking the questions. Though most such databases provide user-friendly web portals to access information, powerful analyses are enabled by programmatic access—computer-friendly mechanisms for accessing large quantities of data that can be parsed by third-party tools that perform tasks not built into existing web applications.

One such mechanism is the "Dimensions on Google BigQuery" dataset, which organises the Dimensions corpus (more than 127 million publications, 6 million grants, 695,000 clinical trials, and so on) ("The data in Dimensions" 2022) into a relational database model hosted on Google's powerful BigQuery platform that enables quick execution of queries that would be too large or too slow on more conventional database platforms such as MySQL. However, information about these entities—and the billions of links between them—is necessarily spread out over 10 tables with dozens of fields each ("Data Source Tables" 2022), presenting a learning curve much steeper than that of the Dimensions web application.

We present an open source Python library that streamlines the process of generating large-scale network visualisations of scientific research related to COVID-19. The library is available on GitHub[1] and relies on the public-domain COVID-19 Dimensions on BigQuery[2] dataset in order to extract and calculate proximity relationships between scientific entities of interest, although it can be easily configured to run on the full Dimensions data. The library includes components for both data extraction and processing, so that it can be consumed by the VOSviewer[3] online visualisation tool. More output visualisations are being developed and will be added over the coming months.

As previously argued (Hook, Porter, 2021), by leveraging cloud-computing infrastructures such as the one provided by Dimensions on Google BigQuery, it is possible to radically transform the way research analytics is done. The main contribution of this work is to provide a reusable open-source tool that practically demonstrates how to leverage such technologies, in order to generate insightful network representations of selected aspects of the scientific research landscape e.g. organisation collaboration networks and topic co-occurrence networks.

**Dataset**

The library by default uses the freely available COVID19 Dimensions dataset (Hook et al., 2020). The dataset contains all COVID-19 related published articles, preprints, datasets, grants and clinical trials from Dimensions free for anyone to access. The data was initially released in early 2021 in CSV format and subsequently as a public-domain BigQuery dataset to help the research community stay up to date and greatly reduce the time that would otherwise be required to collate this information from many disparate sources. The dataset is updated daily and, at the time of writing, contains more than 1.1 million documents (note: we used this dataset to make it easier to reproduce the work presented in this paper, however by modifying the default library settings it is also possible to point to the full Dimensions.ai dataset).

Table 1. Summary of data included in the COVID19 Dimensions Dataset

| Entity | Records |
| --- | --- |
| Publications | 1,031,972 |
| Clinical Trials | 14,723 |
| Grants | 16,703 |
| Patents | 41,473 |
| Datasets | 32,784 |
| Organizations | 36,670 |

---

[1] https://github.com/digital-science/dimensions-network-gen
[2] https://console.cloud.google.com/marketplace/product/digitalscience-public/covid-19-dataset-dimensions
[3] https://www.vosviewer.com/

**Implementation**

The library uses Python for data processing and displays network data using VOSviewer, a software tool for visualising network data that is available across multiple platforms, including a browser-based version (van Eck, 2010).. The data extraction component is generic and outputs a data format that can also be ingested by other visualisation libraries, which we are planning to include in the library in the coming months.

The library can be broken down into three main components: 1) a user-created SQL input query, 2) the BigQuery data extraction & network generation module, 3) the data transformation & visualisation component (Figure 1).

Figure 1: Architecture of *dimensions-network-gen* Python library

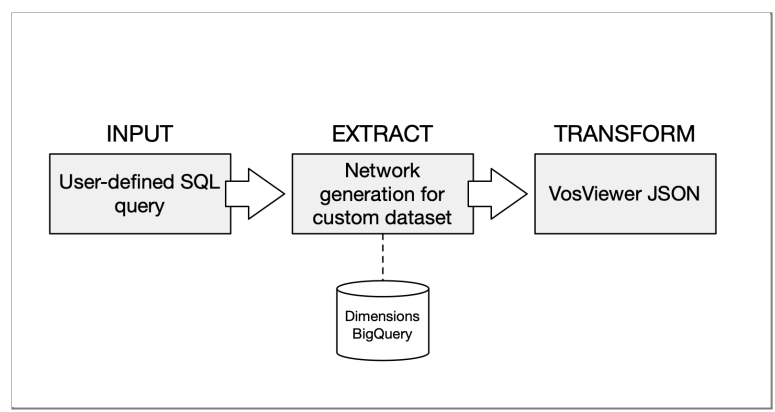

*1. User Input*

It is possible to generate network analyses on the whole COVID19 database, or using a selected subset of data. This is achieved by letting users input any SQL query defining a COVID-19 document subset of interest (e.g. a group of journals, or a group of countries). Users can collect a library of queries of interest by storing them in a folder and then running the extraction script on all of them.

For example, the following query selects only documents added to the database in the last 30 days:

```
select id
from `covid-19-dimensions-ai.data.publications`
where
EXTRACT(DATE FROM date_inserted) >= DATE_ADD(CURRENT_DATE(), INTERVAL -30 DAY)
```

Note how the input query simply returns a set of record IDs—users need only define those fields relevant for their filters; the more complicated task of joining and arranging information about those publications is handled by the library.

*2. Data Extraction*

The data extraction step deals with the extraction of data from BigQuery and calculation of the network representation. Currently we have included two possible network calculations:

1. <u>Concept co-occurrence network.</u> This query generates two-concept pairs and counts how many publications are shared between these concepts (note: concepts in

Dimensions are publication-level keywords normalised and weighted based on a relevancy score).
2. Organisation network. This query generates two-organisations pairs (from the authors affiliations) and counts how many publications are shared between these organisations.

Both the extraction and network calculation steps are achieved using a single SQL query. The query includes the user input query (`user-provided-subquery`) and parameters values for max number of nodes and min weight of edges to be included in the result (`@max_nodes`, `@min_edge_weight`). The gist of the query lies in the double `CROSS JOIN UNNEST`. This mechanism allows to traverse a potentially very large number of relationships in seconds and to expose all relevant combinations of co-authoring organisations within the same data structure.

Figure 2: SQL template query for the collaboration network generation

```
WITH subset AS (
    {user-provided-subquery}
),
top_nodes AS (
    SELECT orgid, COUNT(p.id) AS pubs
    FROM `covid-19-dimensions-ai.data.publications` p
    CROSS JOIN UNNEST(p.research_orgs) orgid
    WHERE id IN (SELECT id FROM subset)
    GROUP BY 1
    ORDER BY 2 DESC
    LIMIT @max_nodes
),
links AS (
    SELECT
    CONCAT(g1.name, ' (', org1_id, ')' ) AS org1
    ,CONCAT(g2.name, ' (', org2_id, ')' ) AS org2
    ,COUNT(DISTINCT p.id) AS collabs
    FROM `covid-19-dimensions-ai.data.publications` p
    CROSS JOIN UNNEST(p.research_orgs) org1_id
    CROSS JOIN UNNEST(p.research_orgs) org2_id
    INNER JOIN `covid-19-dimensions-ai.data.grid` g1 ON org1_id=g1.id
    INNER JOIN `covid-19-dimensions-ai.data.grid` g2 ON org2_id=g2.id
    WHERE
    p.id IN (SELECT id FROM subset)
    AND org1_id > org2_id -- to prevent dupes
    AND org1_id IN (SELECT orgid FROM top_nodes)
    AND org2_id IN (SELECT orgid FROM top_nodes)
    GROUP BY 1,2
)
SELECT *
FROM links
WHERE collabs >= @min_edge_weight
```

*3. Data Transformation & Visualisation*
In this step the data extracted from BigQuery gets converted into a VOSviewer JSON file and packaged up into an HTML application that can be viewed in a browser. The Python library also includes a local server component that can be used to view the files locally on a computer. Example images of the VOSviewer networks generated are included below.

Figure 3: VOSviewer rendering of the organisation network for the *last 30 days* query

Figure 4: VOSviewer rendering of the concepts network for the *last 30 days* query

**Conflict of interest**

The co-authors are current (MP) and former (RJA) employees of Digital Science, the creator and provider of Dimensions.